# MULTI-PATH ROUTE DETERMINATION METHOD FOR NETWORK LOAD BALANCING IN FAP-BASED WSNS USING FUZZY LOGIC

Won-Jin Chung[1] and Tea-Ho Cho[2]

[1]College of Information and Communication Engineering, Sungkyunkwan University, Suwon, Republic of Korea
[2]College of Software, Sungkyunkwan University, Suwon, Republic of Korea

## ABSTRACT

*A flooding attack in wireless sensor networks is a type of threat that shortens the lifetimes of the sensor networks. Although flooding attack prevention techniques have been proposed, if a continuous flooding attack occurs, the sensor node energy is depleted during detection. In this paper, we use multi-path routing to solve this problem. In order to balance the load of the sensor node, energy balancing of the sensor node is controlled by determining the number of pathways using fuzzy logic. By adjusting the energy balancing of the sensor nodes, the number of energy-exhausting sensor nodes can be reduced. As a result, when a flooding attack occurs, the energy efficiency of the sensor node is increased by determining the number of pathways.*

## 1. INTRODUCTION

Wireless sensor networks (WSNs) consist of a number of small sensor nodes that detect various types of information, such as temperature, sound, and vibration. Additionally, a base station (BS) is used to collect information detected by the sensor nodes. In addition, the sensor nodes are deployed randomly over a wide area and autonomously configure their communication. WSNs are used in various fields, including battlefields, smart cities, and intelligent transportation systems. A sensor node can transmit monitored information to the BS through wireless communication, allowing the results to be checked periodically or in real time [1]. However, since WSNs are networks in which sensor nodes are randomly placed and communicate with each other, regions where sensor nodes are arranged in a concentrated manner or where sensor nodes are missing may occur. Furthermore, since sensor nodes use limited computing power and wireless communication, malicious attackers can easily capture and compromise the sensor nodes. Additionally, the sensor nodes have a limited amount of energy and it is difficult to recharge them. When an attacker uses a compromised node for an attack, the sensor node consumes more energy than when a normal packet is dropped due to an attack or when a sensor node transmits and receives a normal packet [2-3]. Therefore, if the residual energy of the sensor node is low, the sensor node energy is exhausted by the attack. If the number of energy-exhausted sensor nodes is large, shadowed areas occur and network life is shortened. Additionally, attackers can use a compromised node to attempt other attacks, such as selective forwarding attacks, sinkhole attacks, and Sybil attacks. Among these types of attacks, flooding attacks shortens the lifetime of the sensor network. To prevent such attacks, flooding attack prevention (FAP) has been proposed [4]. FAP uses the ad-hoc on-demand distance vector (AODV) routing protocol and finds the destination node through the expanding ring search (ERS) algorithm [5]. ERS is an algorithm that finds a destination node through multiple route request (RREQ) and route reply (RREP) retransmissions. The ERS algorithm was proposed to help find the destination node by sending an RREQ from the beginning in a wide range of networks. The ERS algorithm is a method that can be used to expand the range of the search sequentially [6]. Therefore, the destination node is found through a small number of RREQ retransmissions. However, a large amount of energy is consumed in the sensor node during the destination detection process of the AODV protocol. For this reason, the energy of a low-energy sensor node in the detector can be exhausted during the flooding attack detection process. Also, if the number of depleted sensor nodes is large, the sensor network's lifetime is shortened. In this paper, we improve the energy efficiency of the sensor nodes by distributing the load of sensor nodes using multiple pathways and by balancing the sensor nodes. In order to distribute the load of the sensor nodes efficiently, we propose a method that uses a fuzzy system to determine the number paths. Section 2 of this paper discusses flooding attacks, FAP, and multi-path routing. Section 3 describes the proposed method through fuzzy logic, and section 4 includes the experimental results of the proposed method. In the last section, our conclusions and suggested future research are presented.

## 2. RELATED WORKS

### 2.1. Flooding attack

A flooding attack is a type of denial of service attack (DOS) that floods a large number of packets and consumes sensor network resources [7]. When a flooding attack occurs, the sensor network is paralyzed because a sensor node fails to process all of its packets in a short period of time. In addition, the sensor node is affected by an attack that continuously consumes energy by imposing a load on the sensor node in the process of transmitting a large number of packets caused by a flooding attack. A flooding attack in WSNs generates a large number of false packets from a compromised node and transmits these along various paths. Since a sensor node receiving a false packet stores various transmission paths, the storage space of the path table is filled with an incorrect path. This is because, when a new event packet is transmitted, the path of a false packet is filled in the path table; thus, a situation occurs in which the packet cannot be transmitted. Also, since the sensor node transmits the packet received by the BS when the packet is received, the sensor nodes included in the path consume energy continuously while receiving the false packet and transmitting it to the BS. For this reason, a load is generated on the sensor node during the process of transmitting and receiving false packets. Also, since sending a large number of false packets results in continuous energy consumption, a low-energy sensor will undergo energy depletion. If there are many energy-exhausted nodes when an event occurs, a packet may not be transmitted to the BS due to the presence of depleted nodes. Thus, the sensor network's lifetime is shortened.

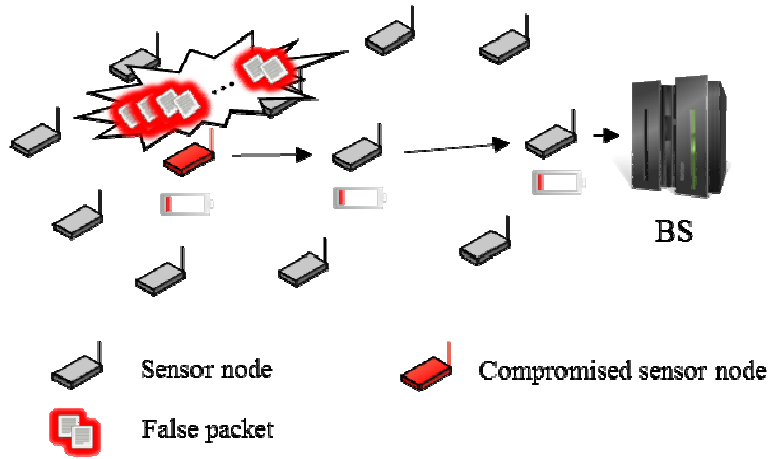

Figure 1. Flooding attack

## 2.2. FAP

Herein, we use the FAP scheme proposed by Y. Ping to prevent flooding attacks from WSNs. This scheme effectively prevents flooding attacks in ad hoc networks. Therefore, the constraints of the sensor nodes should be considered before applying this technique to WSNs. To prevent flooding attacks, each node has two tables. The first table is the blacklist table. When a flooding attack occurs, the blacklist table identifies the flooding attack through the FAP scheme and adds the suspicious node ID of the flooding attack to the blacklist. The second table is the rate-RREQ table. This table has two columns: the node-ID and the RREQ-time. The node-ID includes the neighbor node ID and the RREQ-time records the time when the neighbor node transmits the RREQ packet. After the sensor node has the blacklist table and the rate-RREQ table, the FAP scheme checks the source node ID and the packet generation time at the BS when an attacker attempts a flooding attack. If a large number of packets arrive, and packets with the same node IDs and similar creation times are larger than a preset threshold value, it is determined that a flooding attack is underway. In the BS, a source node ID that is suspected to be the origin of the flooding attack is entered into the blacklist and the updated blacklist is transmitted to all sensor nodes. If the blacklist is transmitted to all of the sensor nodes and the continuous packet is transmitted, the lower node can be compared to the node ID belonging to the blacklist. If this node ID is in the blacklist, it is determined that the attack is a flooding attack and all packets transmitted from the sensor node belonging to the blacklist are blocked. In contrast, if there is no matching node ID in the blacklist, it is treated as a normal packet and forwarded to the BS.

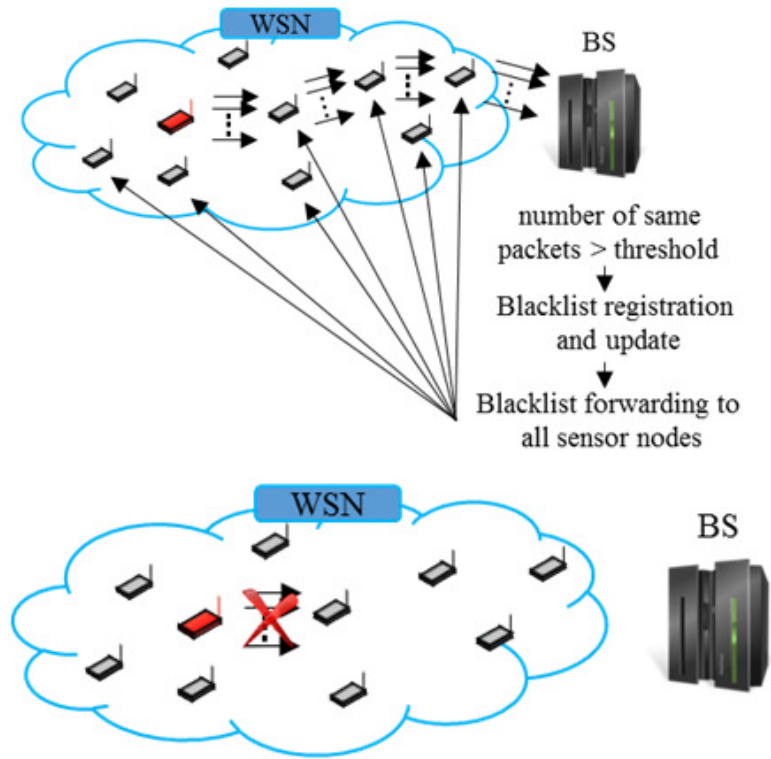

Figure 2. Flooding attack prevention

## 2.3. Multipath routing

The routing of the sensor networks uses multi-hop routing via packet forwarding rather than single-hop routing; this is done because the distribution of sensor nodes is dense. Single-path routing enables rapid routing and data transfer because each sensor node establishes only one path before sending the collected data [8-9]. However, a single path can be rendered inoperable by a defect in the sensor node during packet forwarding. Therefore, the use of multi-path routing provides improved transmission reliability, fault tolerance, congestion control, and quality of service. The multi-path routing scheme establishes multiple pathways, distributes the packets to the established path, and transmits the packets sequentially to balance sensor nodes that are included in the path. The multi-path routing scheme reduces the load on a sensor node and extends the lifespan of a sensor network. However, when multiple paths are set up, the number of hops increases compared to when a packet is transmitted using a single path. In multi-path routing, multiple paths are maintained for data routing. Therefore not only the shortest path but also relatively long distance paths are also available for data routing. Therefore, as the distance increases, the number of hops from the event occurrence area to the BS increases, and data transmission in delayed than using a single shortest path in the data transmission process. Also, by reducing the energy of the sensor nodes periodically, the overall network energy efficiency is reduced.

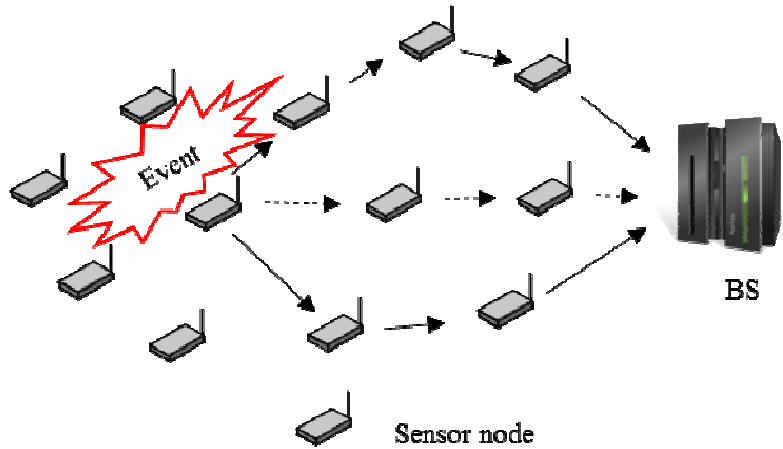

Figure 3. Multi-path routing

## 3. PROPOSED SCHEME

### 3.1. Assumptions

Sensor nodes are placed randomly and the BS either estimates or knows the remaining energy, status, and hop counts for each sensor node. The preset sensor node ID cannot be changed. Flooding attacks occur on compromised nodes.

### 3.2. Fuzzy system

The proposed scheme in this paper uses multi-path routing to disperse the sensor nodes, improve the sensor node energy efficiency if a flooding attack occurs, and reduce the number of exhausted sensor nodes to extend the sensor network lifetime. When the number of multi-paths is set small, the load balancing efficiency is low. On the other hand, if a large number of multi-paths is set, a path in which the number of hops is increased can be selected for data routing. In this case, the sensor network energy efficiency deteriorates. Therefore, the proposed method improves the load balancing efficiency and the sensor network energy efficiency by setting the number of suitable multi-path through the fuzzy system. Therefore, we use fuzzy logic to determine the appropriate number of pathways and distribute a number of false packets (generated by the flooding attack) on as many paths as the determined number of pathways. During the flooding attack detection process of the FAP scheme, the energy depletion of a sensor node that has low residual energy is prevented. The input parameters are the hop count (HC), residual energy (RE), and the number of participating paths in the sensor node (NPP). The output parameter is the number of multiple pathways (NMP).

Input parameter

HC = {VL (VeryLarge), L (Large), M (Medium), S (Small)}

RE = {VL (VeryLarge), L (Large), M (Medium), S (Small), VS (VerySmall)}

NPP = {H (High), M (Medium), L (Low)}

Output parameter

NMP = {VS (VerySmall), S (Small), M (Medium), L (Large), VL (VeryLarge)}

International Journal of Computer Science & Information Technology (IJCSIT) Vol 9, No 6, December 2017

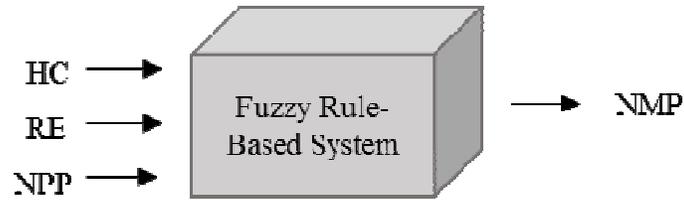

Figure 4. Fuzzy system

The number of hops is an important factor when determining the number of pathways. If multiple paths are used, the number of hops increases relative to when a single path is used. In such cases, energy consumption occurs along the path, influencing the energy efficiency of the sensor network. Additionally, when the residual energy of the sensor node is small, energy is exhausted in the routing setting and the flooding attack detection process should be rerouted. In this case, additional energy consumption by the sensor networks occurs. Finally, if the same sensor node is included in several paths, an increased load will occur on the sensor node. Therefore, in order to balance the load of the sensor node, the number of paths that a sensor node can participate in should be set. Table 1 shows a brief overview of the fuzzy rules.

Table 1. Fuzzy rules

| Rule | Input | | | Output |
|---|---|---|---|---|
| | HC | RE | NPP | (NMP) |
| 0 | VL | VL | H | VS |
| 1 | VL | VL | M | VS |
| 2 | VL | VL | L | S |
| . | . | . | . | . |
| . | . | . | . | . |
| . | . | . | . | . |
| 15 | L | VL | H | VS |
| 16 | L | VL | M | VS |
| 17 | L | VL | L | S |
| 18 | L | L | H | VS |
| 19 | L | L | M | S |
| . | . | . | . | . |
| . | . | . | . | . |
| . | . | . | . | . |
| 31 | L | VS | L | L |
| 32 | M | VL | H | S |
| 33 | M | VL | M | S |
| 34 | M | VL | L | M |
| . | . | . | . | . |

| .  | .  | .  | .  | .  |
|----|----|----|----|----|
| .  | .  | .  | .  | .  |
| 43 | M  | VS | M  | L  |
| 44 | M  | VS | L  | VL |
| 45 | S  | VL | H  | S  |
| 46 | S  | VL | M  | M  |
| .  | .  | .  | .  | .  |
| .  | .  | .  | .  | .  |
| .  | .  | .  | .  | .  |

The following is a fuzzy membership function.

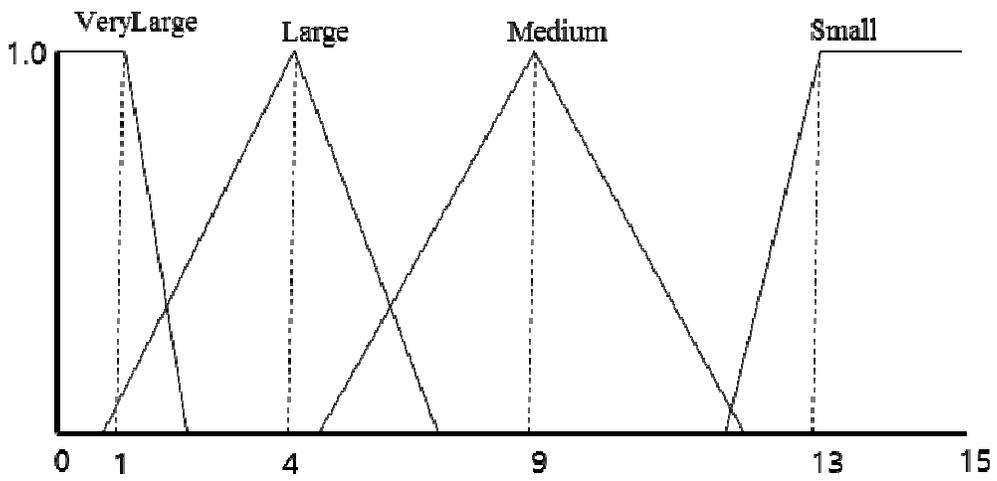

(a) HC

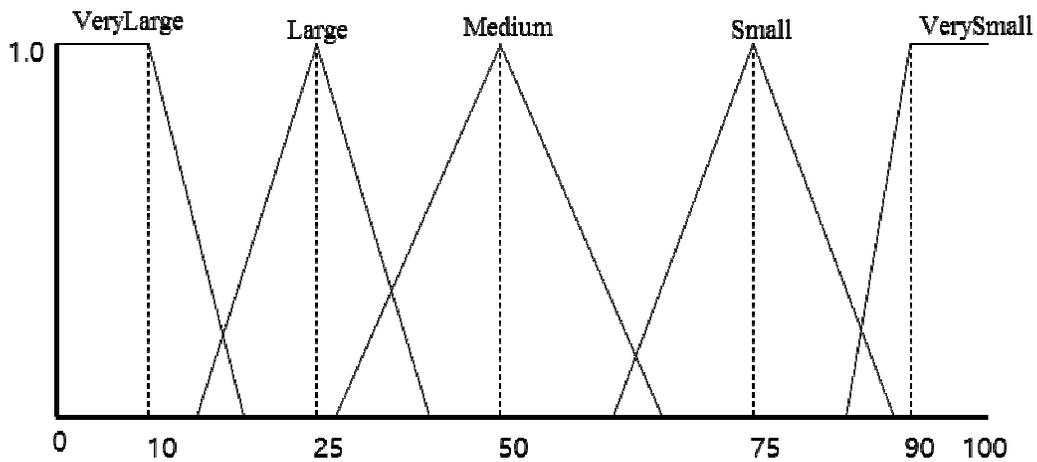

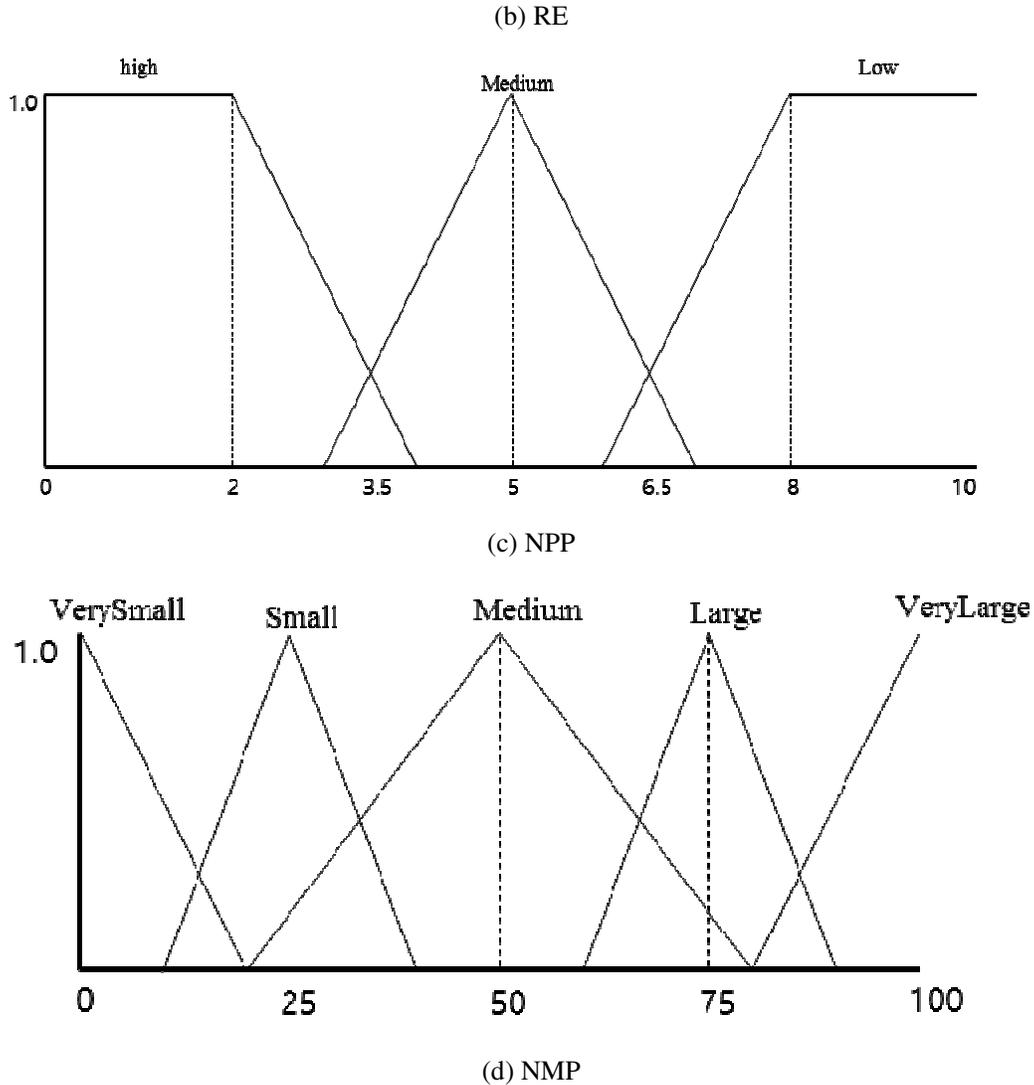

(b) RE

(c) NPP

(d) NMP

Figure 5. Membership function

## 4. EXPERIMENTAL RESULTS

The sensor field size used in this experiment is 1000 × 1000 ($m^2$) and the number of sensor nodes is 500. The type of the sensor node used in this experiment follows the specifications of the mica2 model and is simulated via C++. The initial energy of the mica2 model is powered by two AA batteries. However, since the installation of the sensor node is applied in an older area, there is also energy consumption resulting from energy discharge. Also, energy consumption of the sensor node occurs due to packet transmission that occurred previously. Thus, the energy of the sensor nodes remains below the initial energy. Therefore, the energy of each sensor node is randomly set at a value that is lower than the initial energy (1 J). The false packet sizes used in flooding attacks are set to the maximum packet size used in TinyOS, which is 29 bytes [10]. When transmitting packets from a sensor node to the BS, the energy consumed per byte is 16.25 μJ. When a packet is received, 12.25 μJ per byte is consumed [11].

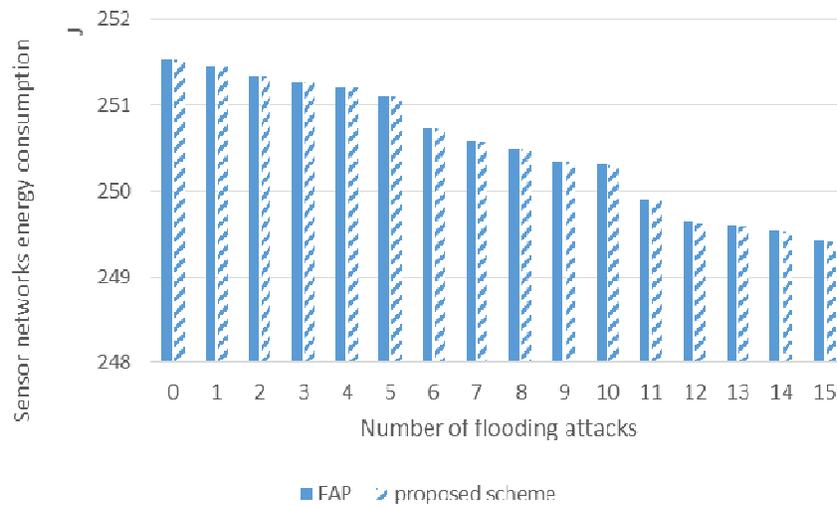

Figure 6. Energy consumption due to flooding attacks in sensor networks

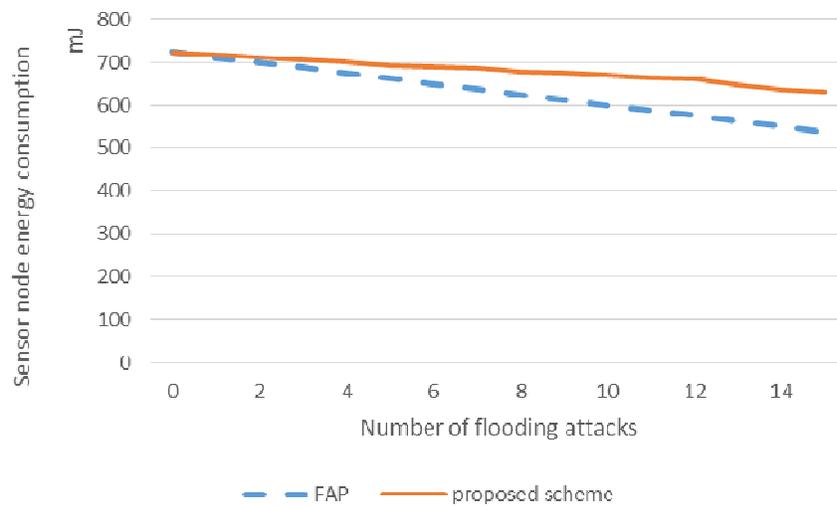

Figure 7. BS energy efficiency of adjacent sensor nodes

Figure 6 shows the sensor network energy decrease due to the number of flooding attacks; a similar energy reduction occurs in the FAP and proposed schemes. In addition, Figure 7 shows the residual energy amount of the sensor node installed around the BS when a flooding attack occurs. It is assumed that the flooding attack occurs 15 times during the experiments. The proposed method and the FAP method are again compared by analyzing the energy of the sensor nodes installed around the BS. The experimental results show that the proposed scheme consumes less energy than the FAP scheme. Thus, if a flooding attack occurs, the proposed scheme shows a sensor node energy efficiency that has been improved by approximately 10% compared to the FAP scheme. Additionally, when the FAP scheme is used, two more sensor nodes (relative to the proposed scheme) in the path experience energy depletion.

## 5. CONCLUSIONS

The FAP scheme can be effectively used to prevent flooding attacks. However, it does not consider the residual energy of the sensor node. As a result, FAP leads to energy depletion in low-energy sensor nodes during the flooding attack detection process. When a sensor field has a depleted sensor node, the sensor network requires a new route reset. Thus, the sensor network consumes additional energy. To address these problems, we used multiple paths to distribute and balance the load on the sensor node. However, the sensor network consumes additional energy due to the fact that the number of hops increases when using multiple pathways instead of a single path. Therefore, we propose a scheme to increase the energy efficiency of sensor nodes by determining the number of paths through fuzzy logic. By determining the appropriate number of multi-paths, we can balance the energy of the sensor nodes and reduce the number of nodes that undergo energy depletion. Additionally, the energy efficiency of the sensor network is improved by limiting the number of hops generated in multi-path routing. Experimental results show that the energy efficiency of the BS peripheral sensor nodes is increased by about 10% compared to the FAP scheme. This reduces the energy depletion of sensor nodes through efficient energy balancing. However, if the attacker changes the source node ID every time a flooding attack occurs, the normal sensor node ID is included in the blacklist, even if the flooding attack is prevented by the FAP method. If the packet is transmitted, the normal sensor node is blocked. If this happens repeatedly, the sensor network can become paralyzed. Therefore, in future research, mutual verification will be conducted to block the change if the sensor node ID is altered.

## ACKNOWLEDGEMENTS

This research was supported by Basic Science Research Program through the National Research Foundation of Korea (NRF) funded by the Ministry of Education, Science and Technology (No. NRF-2015R1D1A1A01059484)

**Authors**

**Won Jin Chung**

Received a B.S. degree in Information Security from Baekseok University in 2016, and is now working toward an M.S. degree in the Department of Electrical and Computer Engineering at Sungkyunkwan University.

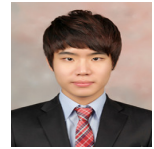

**Tea Ho Cho**

Received a Ph.D. degree in Electrical and Computer Engineering from the University of Arizona, USA, in 1993, and B.S. and M.S.degrees in Electrical and Computer Engineering from Sungkyunkwan University, Republic of Korea, and the University of Alabama, USA, respectively. He is currently a Professor in the College of Information and Communication Engineering, Sungkyunkwan university korea.

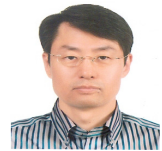